\documentclass[preprint,pra,showpacs,nofootinbib]{revtex4}
\usepackage{mathrsfs}
\usepackage{float,epsfig}
\usepackage{dcolumn}% Align table columns on decimal point
\usepackage{bm}% bold math
\usepackage{graphicx}% Include figure files
\usepackage{dcolumn}% Align table columns on decimal point
\usepackage{bm}% bold math
\usepackage{amsmath,amssymb,amsthm}
\usepackage[colorlinks=true,linkcolor=blue]{hyperref}
\begin{document}
\title{\bf Understanding  Hawking radiation in \\ the framework of open quantum systems}
\author{ Hongwei Yu and Jialin Zhang
%\footnote{Corresponding author}
}
\affiliation{Department of Physics and Institute of  Physics,\\
Hunan Normal University, Changsha, Hunan 410081, China}

\date{\today}

\begin{abstract}

We study the Hawking radiation in the framework of open quantum
systems by examining the time evolution of a detector (modelled by
a two-level atom) interacting with vacuum massless scalar fields.
The dynamics of the detector is governed by a master equation
obtained by tracing over the field degrees of freedom from the
complete system. The nonunitary effects are studied by analyzing
the time behavior of a particular observable of the detector,
i.e., its admissible state, in the Unruh, Hartle-Hawking, as well
as Boulware vacua outside a Schwarzschild black hole. We find that
the detector in  both the Unruh and Hartle-Hawking vacua would
spontaneously excite with  a nonvanishing probability the same as
what one would obtain if there is thermal radiation at the Hawking
temperature from the black hole, thus reproducing the basic
results concerning the Hawking effect in the framework of open
quantum systems.

\end{abstract}
\pacs{04.70.Dy, 03.65.Yz, 04.62.+v} \maketitle
\section{Introduction}

Black holes are intriguing objects and are worth studying in all
possible varieties. The idea of black holes has been proven to be
highly fruitful. In particular, Hawking's discovery that black holes
are not, after all, completely black, but quantum mechanically, emit
radiation with a thermal spectrum~\cite{Hawking:rv},  has provided
us with the understanding that black holes may play the role of
``Rosetta stone" to relate gravity, quantum theory and
thermodynamics. Therefore, Hawking radiation, as one of the most
striking effects that arise from the combination of quantum theory
and general relativity, has attracted widespread interest in physics
community and it has been extensively examined from different
prospectives, yielding different derivations of it. These
derivations include (but not limited to) Hawking's original one
which calculates the Bogoliubov coefficients between the quantum
scalar field modes of the in vacuum states and those of the out
vacuum~\cite{Hawking:rv,Hawking:sw}, an Euclidean quantum gravity
derivation~\cite{Gibbons:1976ue} which has been interpreted as a
calculation of tunnelling through classically forbidden
trajectory~\cite{Parikh:1999mf}, an approach based upon string
theory~\cite{SC,Peet}, an interesting proposal which ties its
existence to the cancellation of gravitational anomalies at the
horizon~\cite{Robinson:2005pd}, and a recent study which reveals an
interesting relationship between the existence of Hawking radiation
and the spontaneous excitation of
 atoms using the DDC formalism~\cite{ddc} that separates the contributions of vacuum fluctuations and radiation reaction to
the rate of change of the mean atomic energy~\cite{yu1}.

In the current paper, we shall try to understand the Hawking
radiation by examining the time behavior of a static detector
(modelled by a two-level atom) outside a Schwarzschild black hole
immersed in vacuum massless scalar fields using the well-known
techniques developed in the study of open quantum systems. As for
any open system, the full dynamics of the detector can be obtained
from the complete time evolution describing the total system
(detector plus external fields) by integrating over the field
degrees of freedom, which are in fact not observed. It is worth
noting here that an examination of a similar issue, i.e.,  the
Unruh effect associated with uniformly accelerated atoms in the
paradigm of open quantum system has been already been carried
out~\cite{Benatti1}.

The paper is organized as follows. In next Section, we shall
review the basic formalism, the derivation of the master equation
describing system of  the detector plus external vacuum scalar
fields in weak coupling limit and the reduced dynamics it
generates for the finite time evolution of the detector. In
Section III, we apply the method and results of the preceding
Section to discuss the probability of spontaneous transition of
detector from the ground state to the excited states outside a
Schwarzschild black hole. Finally, we conclude with some
discussions in Section IV

\section{the Master Equation}

We shall consider the evolution in the proper time  of a static
detector (two-level atom) interacting with vacuum massless scalar
fields outside a Schwarzschild black hole and assume the combined
system (detector + external vacuum fields) to be initially prepared
in a factorized state, with the detector held static in the exterior
region of the black hole and the fields in their vacuum states. Our
derivation of the master equation in this Section follows closely to
that in Ref.~\cite{Benatti1}. The atom is assumed to be fully
described in terms of a two-dimensional Hilbert space, so that its
states can be represented by a $2\times2$ density matrix $\rho$,
which is Hermitian $\rho^+=\rho$, and normalized ${\rm
{Tr}}(\rho)=1$ with ${\rm {det}}(\rho)\geq0\;.$ In order to achieve
a rigorous, mathematically sound, derivation of the reduced dynamics
of the detector, we shall assume that the interaction between the
detector and the scalar fields are weak so that the finite-time
evolution describing the dynamics of the detector takes the form of
a one-parameter semigroup of completely positive
maps~\cite{Davies,BP}.
 Without loss of generality, we take the total
Hamiltonian for the complete system  to have the form
\begin{equation}
\label{H}
 H=H_s+H_\phi+ \lambda\;H'\; .
 \end{equation}
 Here $H_s$ is the Hamiltonian of
the atom, which in the most generic case takes the form
 \begin{equation}
 H_s={{\omega_0}\over 2}\sum_{i=1}^3n_i\sigma_i\;,
 \end{equation}
where $\sigma_i\; (i=1,2,3)$ are the Pauli matrices, $\omega_0$
the energy level spacing and $\mathbf{n} =(n_1,n_2,n_3)$  a unit
vector. In the present paper, we will take, for simplicity,
$\mathbf{n}$  to be along the third axis such that $H_s$
simplifies to
\begin{equation}
H_s={\omega_0\over 2}\sigma_3 \;.
\end{equation}
 $H_\phi$ is the standard Hamiltonian
of massless, free scalar fields, details of which need not be
specified here and $H'$ is the interaction Hamiltonian of the atom
with the external scalar fields and is assumed to be given by
\begin{equation}
H'= \sigma_{3}\Phi(x)\;.
\end{equation}
%%%%%%%%%%%%%%%%%%%%%%%%%%%%%%%%%%%%%%%%
 It
should be pointed out that the coupling constant $\lambda$ in
(\ref{H}) is small, and this is required by our assumption that
the interaction of the atom with the scalar fields is weak.

 Initially,  the complete system is described by the total density
 $\rho_{tot}=\rho(0)\otimes|0\rangle\langle0|\;,$
where $\rho(0)$ is the initial reduced density matrix of the atom,
and $|0\rangle$ is vacuum state of the field $\Phi(x)$.
  In the frame of the atom,  the evolution in the proper time $\tau $ of the total density
  matrix,
$\rho_{tot}$,  of the complete system satisfies
\begin{equation}
\frac{\partial\rho_{tot}(\tau)}{\partial\tau}=-iL_H[\rho_{tot}(\tau)]\;,
\end{equation}
where the symbol $L_H$ represents the Liouville operator
associated with $H$
\begin{equation}
L_H[S]\equiv[H,S]\;.
\end{equation}
The dynamics of the atom can be obtained by tracing over the field
degrees of freedom, i. e., by applying the trace projection to the
total density matrix $\rho(\tau)={\rm
{Tr}}_{\Phi}[\rho_{tot}(\tau)]\;$.

 In the limit of weak-coupling which we assume in the present paper, the reduced density  is found to obey
an equation in the Kossakowski-Lindblad
form~\cite{Lindblad,Benatti0}
\begin{equation}
\frac{\partial\rho({\tau})}{ \partial {{\tau}}}= -i \big[H_{\rm
eff},\, \rho({\tau})\big]
 + {\cal L}[\rho({\tau})]\ ,
\label{master}
\end{equation}
where
\begin{equation}\label{lindblad}
{\cal L}[\rho]=\frac{1}{2} \sum_{i,j=1}^{3} a_{ij}\big[2\,
\sigma_j\rho\,\sigma_i -\sigma_i\sigma_j\, \rho
-\rho\,\sigma_i\sigma_j\big]\;.
\end{equation}
The  matrix $a_{ij}$ and the effective Hamiltonian $H_{\rm eff}$
are determined by the Fourier  and Hilbert transforms of the field
vacuum correlation functions
\begin{equation}
G^+(x-y)=\langle0|\Phi(x)\Phi(y)|0 \rangle\label{twogreen}\;,
\end{equation}
which are defined as
\begin{equation}
{\cal G}(\lambda)=\int_{-\infty}^{\infty} d\tau \,
e^{i{\lambda}\tau}\, G^{+}\big(x(\tau)\big)\; , \label{fourierG}
\end{equation}
\begin{equation}
{\cal K}(\lambda)= \frac{P}{\pi i}\int_{-\infty}^{\infty} d\omega\
\frac{ {\cal G}(\omega) }{\omega-\lambda} \;. \label{kij}
\end{equation}
Then  the coefficients of the Kossakowski matrix $a_{ij}$ can be
written explicitly as
\begin{equation}
a_{ij}=A\delta_{ij}-iB
\epsilon_{ijk}\delta_{k3}+C\delta_{i3}\delta_{j3}
\end{equation}
with
\begin{equation} \label{ABC}
A=\frac{1}{2}[{\cal {G}}(\omega_0)+{\cal{G}}(-\omega_0)]\;,\;~~~
B=\frac{1}{2}[{\cal {G}}(\omega_0)-{\cal{G}}(-\omega_0)]\;,~~~~
C={\cal{G}}(0)-A\;.
\end{equation}
Meanwhile, the effective Hamiltonian,  $H_{\rm eff}$,  contains a
correction term, the so-called Lamb shift, and one can show that
it can be obtained by replacing $\omega_0$ in $H_s$ with a
renormalized energy level spacing $\Omega$ as
follows~\cite{Benatti1}
\begin{equation}
H_{\rm eff}=\frac{\Omega}{2}\sigma_3=\{\omega_0+i[{\cal
K}(-\omega_0)-{\cal K}(\omega_0)]\}\sigma_3\;,
\end{equation}
where a suitable substraction is assumed in the definition of
${\cal K}(-\omega_0)-{\cal K}(\omega_0)$ to remove the logarithmic
divergence which would otherwise be present.

To facilitate the discussion of the properties of solutions for
(\ref{master}) and (\ref{lindblad}), let us express the density
matrix $\rho$ in terms of the Pauli matrices,
\begin{equation}\label{rho-b}
\rho({\tau})=\frac{1}{2}\bigg(1+\sum_{i=1}^{3}\rho_i({\tau})\sigma_i\bigg)\;.
\end{equation}
Substituting Eq.~(\ref{rho-b}) into Eq.~(\ref{lindblad}), it is
easy to show that the Bloch vector $|\rho({\tau})\rangle$ of
components $\{\rho_1({\tau}),\rho_2({\tau}),\rho_3({\tau})\}$
satisfies
\begin{equation}\label{sh-eq}
\frac{\partial}{\partial{{\tau}}}|\rho({\tau})\rangle=-2{\cal{H}}|\rho({\tau})\rangle+|\eta\rangle\;,
\end{equation}
where $|\eta\rangle$ denotes a constant  vector $\{0,0,-4B\}$. The
exact form of the matrix ${\cal{H}}$ reads
\begin{equation}
{\cal{H}}=\left(
\begin{array}{ccc}
2A+C& \Omega/2& 0\\ -\Omega/2&2A+C&0 \\ 0&0&2A
\end{array}\right)\;.
\end{equation}
Eq.~(\ref{sh-eq}) can be solved formally, with the solution being
written as
\begin{equation}\label{rhot}
|\rho(\tau)\rangle=e^{-2{\cal{H}}{\tau}}|\rho(0)\rangle+(1-e^{-2{\cal{H}}{\tau}})|\rho_\infty\rangle\;,
\end{equation}
with
 \begin{equation}\label{beta1}
|\rho_\infty\rangle=\frac{1}{2}{\cal{H}}^{-1}|\eta\rangle=-\frac{B}{A}\left(
\begin{array}{c}
 0\\ 0 \\ 1
\end{array}\right)\;.
\end{equation}
 Here the matrix (operator) $e^{-2{\cal{H}}{\tau}}$  is
defined by series expansion as usual. However, since ${\cal{H}}$
obeys a cubic eigenvalue equation, so powers of ${\cal{H}}$ higher
than 2 can always be written in terms of  combinations of
${\cal{H}}^2$,\;${\cal{H}}$ and $I$, the $3\times 3$ unit matrix.
Actually,  three eigenvalues of ${\cal{H}}$ can be found
explicitly: $\lambda_1=2A,\; \lambda_{\pm}=(2A+C)\pm i \Omega /2$.
One can then show that
\begin{equation}\label{expfunction}
e^{-2{\cal{H}}\tau}=\frac{4}{\Omega^2+4C^2}\bigg\{e^{-4A{\tau}}\Lambda_1+2e^{-2(2A+C){\tau}}\bigg[\Lambda_2
\cos(\Omega{{\tau}})+\Lambda_3\frac{\sin(\Omega{{\tau}})}{\Omega}\bigg]\bigg\}\;,
\end{equation}
where
\begin{eqnarray}\label{matrix123}
&&\Lambda_1=\big[(2A+C)^2+\frac{\Omega^2}{4}\big]I-2(2A+C){\cal{H}}+{\cal{H}}^2\;,\nonumber\\
&&\Lambda_2=-2A(A+C)I+(2A+C){\cal{H}}-\frac{1}{2}{\cal{H}}^2\;,\nonumber\\
&&\Lambda_3=2A\big[\frac{\Omega^2}{4}-C(2A+C)\big]I+\big[C(4A+C)-\frac{\Omega^2}{4}\big]{\cal{H}}-C{\cal{H}}^2\;.
\end{eqnarray}
Eq.~(\ref{expfunction}) reveals the effects of decoherence and
dissipation on the atom characterized by the exponentially
decaying factors involving the real parts of the eigenvalues of
${\cal{H}}$ and oscillating terms associated with the imaginary
part. These nonunitary effects can be analyzed by examining the
evolution behavior in time of suitable atom observables. For any
observable of the atom represented by a Hermitian operator
${\cal{O}}$, the time behavior of its mean value is determined by
 \begin{equation}
 \label{meanvalue}
 \langle{\cal{O}}({\tau})\rangle={\rm {Tr}}[{\cal{O}}\rho({\tau})]\;
 \end{equation}
 Let the observable ${\cal{O}}$ be an admissible atom state
 $\rho_f$, then Eq.~(\ref{meanvalue}) yields the probability, ${\cal{P}}_{i\rightarrow{f}}({\tau})$,
  that the atom evolves to the expected state represented by density matrix $\rho_f(\tau)$
from an initial one $\rho(0)\equiv \rho_i$
\begin{equation}\label{prob}
{\cal{P}}_{i\rightarrow{f}}({\tau})={\rm
{Tr}}[\rho_f\;\rho({\tau})]\;.
\end{equation}
If initially the atom is in the ground state, so that its Bloch
vector $|\rho(0)\rangle$ is  $\{0,0,-1\}$, and the final state
$\rho_f$ is the excited state given by the Bloch vector
$|\rho_f\rangle =\{0,0,1\}$, then one has, by making use of
Eqs.~(\ref{rhot}-\ref{prob}),
\begin{eqnarray}\label{pif}
{\cal{P}}_{i\rightarrow{f}}({\tau}) =\frac{1}{2}\bigg(1-e^{-4A
{\tau}}\bigg)\bigg(1-\frac{B}{A}\bigg)\;.
\end{eqnarray}
The probability per unit time of the transition from the ground
state to the excited state, in the limit of infinitely slow
switching on and off the atom-field interaction,  i.e., the
spontaneous excitation rate can be calculated by taking the time
derivative of ${\cal{P}}_{i\rightarrow{f}}({\tau})$ at ${\tau}=0$,
\begin{equation}\label{gammaif}
{\Gamma_{i\rightarrow{f}}}=
\frac{\partial}{\partial{{\tau}}}{\cal{P}}_{i\rightarrow{f}}({\tau})\bigg|_{{\tau}=0}=2A-2B=2{\cal{G}}(-\omega_0)\;.
\end{equation}

\section{Probability of spontaneous transition of the detector outside a Schwarzschild black hole}

Now, we apply the open quantum system formalism developed in the
preceding Section to address the issue of finite time evolution of
a static detector (two-level atom) interacting with vacuum scalar
fields outside a spherically symmetric black hole and calculate
the probability of spontaneous transition and the spontaneous
excitation rate of the detector from the ground state to the
excited state. The metric of a spherically symmetric black hole
can be written, in the Schwarzschild coordinates, as
\begin{equation}
ds^2=\bigg(1-\frac{2M}{r}\bigg)dt^2-\frac{dr^2}{1-2M/r}-r^2(d\theta^2+\sin^2\theta{d\phi^2})\;.
\end{equation}
However, a delicate issue that we have to address first before
moving on is  how the vacuum state of the quantum fields is
determined. Normally, a vacuum state is associated with
non-occupation of positive frequency modes. However, the positive
frequency of field modes are defined with respect to the time
coordinate. Therefore, to define positive frequency, one has to
first specify a definition of time. In a spherically symmetric
black hole background, one definition is the Schwarzschild time,
$t$. The vacuum state, defined by requiring normal modes to be
positive frequency with respect to the Killing vector $\partial/
\partial t$ with respect to which the exterior region is static, is
called the Boulware vacuum~\cite{Boulware}. Other possibilities
that have been proposed are the Unruh vacuum~\cite{Unruh} and the
Hartle-Hawking vacuum~\cite{Hartle-Hawking}. The Unruh vacuum is
defined by taking modes that are incoming from $\mathscr{J}^-$ to
be positive frequency with respect to $\partial/ \partial t$,
while those that emanate from the past horizon are taken to be
positive frequency with respect to the Kruskal coordinate $\bar
u$, the canonical affine parameter on the past horizon.  The
Hartle-Hawking vacuum, on the other hand, is defined by taking the
incoming modes to be positive frequency with respect to $\bar v$,
the canonical affine parameter on the future horizon, and outgoing
modes to be positive frequency with respect to $\bar u$.
Therefore, in what follows, we shall calculate the probability of
spontaneous transition of a static detector from the ground state
to its excited state and spontaneous excitation rate in these
three vacuum states.

 \subsection{The Unruh vacuum}
 We start with the Unruh vacuum.
 The Wightman function for massless scalar fields in the Unruh vacuum can be computed as follows~\cite{wightman1,wightman2,wightman3}
 \begin{equation}
{G_{U}}^+(x,x')=
\sum_{ml}\int^{\infty}_{-\infty}\frac{e^{-i\omega\Delta{t}}}{4
\pi\omega}|\,Y_{lm}(\theta,\phi)\,|^2\bigg[\frac{|\,\overrightarrow{R_l}(\omega,r)\,|^2}{1-e^{-2\pi\omega/\kappa}}
+\theta(\omega)|\,\overleftarrow{R_l}(\omega,r)\,|^2\bigg]d\omega\;,
 \end{equation}
where $\;\kappa=1/(4M)\;$ is the surface gravity of the black
hole.
 The Fourier transform with respect to the proper time, which is related with the coordinate time through,
 \begin{equation}
 d\tau=\sqrt{g_{00}}\,dt=\sqrt{1-{2M\over r}}\,dt\;,
 \end{equation}
  reads
 \begin{eqnarray}\label{Guf}
{\cal {G}}_U(\lambda)&=&\int^{\infty}_{-\infty}e^{i{\lambda}\tau}{G_{U}}^+(x,x')d\tau\nonumber\\
&=&\frac{1}{8\pi{\lambda}}\sum_{l=0}^{\infty}\bigg[\theta({\lambda}\sqrt{g_{00}})(1+2l)|\,\overleftarrow{R_l}({\lambda}\sqrt{g_{00}},r)\,|^2
+\frac{(1+2l)|\,\overrightarrow{R_l}({\lambda}\sqrt{g_{00}},r)\,|^2}{1-e^{-2\pi{\lambda}\sqrt{g_{00}}/\kappa}}\bigg]\;,
\end{eqnarray}
where we have used the relation
\begin{equation} \label{asymp1}
 \sum^l_{m=-l}|\,Y_{lm}(\,\theta,\phi\,)\,|^2= {2l+1 \over
 4\pi}\;.
 \end{equation}
 The summation in Eq.~(\ref{Guf}) is not easy to evaluate
 into a closed form in general. However, its behavior both close to
 the event horizon and in the spatial asymptotic region, which, fortunately, are regions we are most interested in, can be
 found using the following properties of the radial functions
\begin{equation} \label{asymp2}
\sum_{l=0}^\infty\,(2l+1)\,|\overrightarrow{R}_l(\,\omega,r\,)\,|^2\sim\left\{
                    \begin{aligned}
                 &\frac{4\omega^2}{1-\frac{2M}{r}}\;,\;\;\;\quad\quad\quad\quad\quad\quad\quad r\rightarrow2M\;,\cr
                  &\frac{1}{r^2}
\sum_{l=0}^\infty(2l+1)\,|\,{B}_l\,(\omega)\,|^2\;,\quad\;r\rightarrow\infty
                  \;,
                          \end{aligned} \right.
\end{equation}

\begin{equation} \label{asymp3}
\sum_{l=0}^\infty\,(2l+1)\,|\overleftarrow{R}_l(\,\omega,r\,)\,|^2\sim\left\{
                    \begin{aligned}
                 &\frac{1}{4M^2}\sum_{l=0}^\infty(2l+1)\,|\,{B}_l\,(\omega)\,|^2,\quad\;r\rightarrow2M\;,\cr
                  &4\omega^2,\;\;\;\;\quad\quad\quad\quad\quad\quad\quad\quad\quad\quad r\rightarrow\infty
                  \;,\cr
                          \end{aligned} \right.
\end{equation}
and
\begin{equation} \label{asymp4}
\sum_{l=0}^\infty\frac{B_l(\omega)}{\omega}\bigg|_{\omega\rightarrow0}=4M\delta_{l0}\;.
\end{equation}
With the above expression, it is easy to show that
\begin{equation} \mathcal{G}_U(\omega_0)\approx\left\{
\begin{aligned}
 &\frac{1}{8\pi{\omega_0}}\bigg[\frac{4{\omega_0}^2}{1-e^{-2\pi{\omega_0}/\kappa_r}}+
  \frac{\theta({\omega_0})}{4M^2}\sum_{l=0}^\infty(1+2l)|\,B_l({\omega_0\sqrt{g_{00}}})\,|^2\bigg]\;,&r\rightarrow2M\;,
\\
&\frac{1+\coth\big(\pi\omega_0/\kappa_r\big)}{16\pi{\omega_0}{r^2}}
\sum_{l=0}^\infty(1+2l)|\,B_l({\omega_0}\sqrt{g_{00}})\,|^2 +
 \frac{\omega_0g_{00}\theta({\omega_0})}{2\pi}\;,&r\rightarrow\infty\;,
\end{aligned} \right.
\end{equation}
where  $\kappa_r=\kappa/{\sqrt{1-2M/r}}\;$. This leads to the
following behaviors of the coefficients of the Kossakowski matrix
\begin{eqnarray}\label{xishu-u1}
r\rightarrow2M:\left\{ \begin{aligned}
 &A_U\approx
 \frac{1}{64M^2\pi{\omega_0}}\sum_{l=0}^\infty(1+2l)|\,B_l({\omega_0}\sqrt{g_{00}})\,|^2+
 \frac{{\omega_0}}{4\pi}\coth\big(\frac{\pi\omega_0}{\kappa_r}\big)\;,\\
 &B_U\approx
 \frac{1}{64M^2\pi{\omega_0}}\sum_{l=0}^\infty(1+2l)|\,B_l({\omega_0\sqrt{g_{00}}})\,|^2+
 \frac{{\omega_0}}{4\pi}\;,\\
 &C_U=\mathcal{G}_U(0)-A_U\;.
 \end{aligned} \right.
 \end{eqnarray}

\begin{eqnarray}\label{xishu-u2}
r\rightarrow\infty:\left\{ \begin{aligned}
 A_U&\approx
 \frac{\omega_0}{4\pi}+\frac{F^{(+)}(\omega_0)}{r^2}\;,\\
 B_U&\approx\frac{\omega_0}{4\pi}+\frac{F^{(-)}(\omega_0)}{r^2}\;,\\
 C_U&=\mathcal{G}_U(0)-A_U\;,&\;
 \end{aligned} \right.
 \end{eqnarray}
 where the auxiliary function is defined as
 \begin{equation}
 F^{(\pm)}(\omega_0)=
\frac{-1+\coth\big(\frac{\pi{\omega_0}}{\kappa}\big)}{32\pi{\omega_0}}
 \sum_{l=0}^\infty(1+2l)\bigg[
 e^{2\pi{\omega_0}/\kappa}
|\,B_l({\omega_0})\,|^2
 \pm|\,B_l(-{\omega_0})\,|^2
 \bigg]\;.
\end{equation}
%%%%%%%%%%%%%%%%%%%%%%%%%%%%%
 Substitute the above coefficients into Eq.~(\ref{pif}) and
Eq.~(\ref{gammaif}), then the probability of spontaneous
transition to the excited state of the ground state detector and
the spontaneous excitation rate can be obtained respectively
\begin{equation}\label{pu1}
\mathcal{P}^U_{i\rightarrow{f}}({\tau})\approx\left\{
\begin{aligned}
  &\bigg[1-e^{-\frac{\tau\omega_0}{\pi}\coth\big(\pi\omega_0/\kappa_r\big)}\bigg]
\bigg[\frac{1}{1+e^{2\pi\omega_0/\kappa_r}}+\frac{g_{00}(e^{2\pi\omega_0/\kappa_r}-1)}
{2(1+e^{2\pi\omega_0/\kappa_r})}
  \bigg]\;,&r\rightarrow2M\;,
\\
&\frac{1-e^{-\tau\omega_0/\pi}}{e^{2\pi{\omega_0}/\kappa}-1}
\sum_{l=0}^\infty\frac{(1+2l)|\,B_l(-{\omega_0}\sqrt{g_{00}})\,|^2}{4{\omega_0}^2r^2}
\;,&r\rightarrow\infty\;,
\end{aligned} \right.
\end{equation}
and
 \begin{equation}\label{gu1}
\Gamma_{i\rightarrow{f}}^U\approx\left\{
\begin{aligned}
  &\frac{{\omega_0}}{\pi(e^{2\pi{\omega_0}/\kappa_r}-1)}\;,&r\rightarrow2M\;,
\\
&\frac{{f(-\omega_0,r)\,\omega_0}}{\pi(e^{2\pi{\omega_0}/\kappa}-1)}
\;,&r\rightarrow\infty\;,
\end{aligned} \right.
\end{equation}
where
 \begin{equation}
 f(\omega_0,r)=\sum_{l=0}^\infty\frac{(1+2l)|\,B_l({\omega_0}\sqrt{g_{00}})\,|^2}{4{\omega_0}^2r^2}\;.
 \end{equation}
 The above results reveal that, close to the horizon, the
ground state detector in the vacuum would spontaneously excite
with an excitation rate same as what one would expect if there is
a flux of thermal radiation at the temperature
$T=\kappa_r/(2\pi)$. However, as the atom is placed away from the
horizon, this thermal flux is backscattered by the spacetime
curvature, resulting in depletion of part the flux, the effect of
which is described by the greybody factor $f(-\omega_0,r)$. The
depletion is dependent on the atom's radial distance from the
black hole and becomes greater as the detector is placed farther
away. The effective temperature $T$, which can be understood as a
result of combined effects of thermal radiation from the black
hole and the Unruh effect which results from the acceleration with
respect to the local free-falling inertial frame needed to hold
the detector static at a finite radial distance, approaches the
Hawking temperature $T_H=\kappa/(2\pi)\;$ in the spatial
asymptotic region, suggesting the temperature of the thermal
radiation emanating from the horizon is just the Hawking value,
since the acceleration needed to hold the detector static vanishes
in the spatial infinity. This understanding we gain for the Unruh
vacuum in the paradigm of open quantum systems is consistent with
what we have in other different
contexts~\cite{wightman3,green1,green2}.
%%%%%%%%%%%%%%%%%%%%%%%%%%%%%%%%%%%%%%%

%%%%%%%%%%%%%%%%%%%%%%%%%%%%%%%%%%%%%%%%%%%%%%%%%%
\subsection{The Hartle-Hawking vacuum}
Now, let us turn to the Hartle-Hawking vacuum case.  Then the
Wightman function for the scalar field
becomes~\cite{wightman1,wightman2,wightman3}
\begin{eqnarray}
{G_{H}}^+(x,x')&=&\sum_{ml}\int^{\infty}_{-\infty}\frac{|\,Y_{lm}(\theta,\phi)\,|^2}{4
\pi\omega}\bigg[\frac{e^{-i\omega{\Delta{t}}}}{1-e^{-2\pi\omega/\kappa}}|\,\overrightarrow{R_l}(\omega,r)\,|^2
\nonumber\\
&&+\frac{
e^{i\omega\Delta{t}}}{e^{2\pi\omega/\kappa}-1}|\,\overleftarrow{R_l}(\omega,r)\,|^2\bigg]d\omega
\end{eqnarray}
The Fourier transform is given by
\begin{eqnarray}
{\cal{G}}_H(\lambda)&=&\int^{\infty}_{-\infty}e^{i{\lambda}\tau}{G_{H}}^+(x,x')d\tau
\nonumber\\&=&\sum_{l=0}^{\infty}\frac{(1+2l)}{8\pi{{\lambda}}}\bigg[
\frac{|\,\overrightarrow{R_l}({\lambda}\sqrt{g_{00}},r)\,|^2}{1-e^{-2\pi{{\lambda}}/\kappa_r}}+
\frac{|\,\overleftarrow{R_l}(-{\lambda}\sqrt{g_{00}},r)\,|^2}{1-e^{-2\pi{{\lambda}}/\kappa_r}}\bigg]\;,
\end{eqnarray}
 which can be evaluated approximately at the event horizon and in the
 spatial asymptotic region to get
 \begin{equation}
\mathcal{G}_H(\omega_0)\approx\left\{ \begin{aligned}
  &\frac{1+\coth\big(\frac{\pi \omega_0}{\kappa_r}\big)}{16\pi{\omega_0}}\bigg[4{\omega_0}^2+
  \sum_{l=0}^\infty\frac{(1+2l)|\,B_l(-{\omega_0}\sqrt{g_{00}})\,|^2}{4M^2}\bigg]\;,&r\rightarrow2M\;,
\\
&\frac{1+\coth\big(\frac{\pi
\omega_0}{\kappa}\big)}{16\pi{\omega_0}}\bigg[\sum_{l=0}^\infty\frac{(1+2l)|\,B_l({\omega_0}\sqrt{g_{00}})\,|^2}{r^2}+
 4{\omega_0}^2g_{00}\bigg]\;,&r\rightarrow\infty\;.
\end{aligned} \right.
\end{equation}

The corresponding coefficients for $a_{ij}$ then read
\begin{eqnarray}\label{xishu-h1}
r\rightarrow{2M}:~\left\{ \begin{aligned}
 A_H&\approx \frac{\omega_0\coth\big(\frac{\pi\omega_0}{\kappa_r}\big)}{4\pi}+
  \frac{-1+\coth\big(\frac{\pi{\omega_0}}{\kappa_r}\big)}{128M^2\pi{\omega_0}}
\sum_{l=0}^\infty(1+2l)\bigg[|\,B_l({\omega_0}\sqrt{g_{00}})\,|^2
\\&+e^{2\pi{\omega_0}/\kappa_r}|\,B_l(-{\omega_0}\sqrt{g_{00}})\,|^2
\bigg]
\;,\\
 B_H&\approx
\frac{\omega_0}{4\pi}+
  \frac{-1+\coth\big(\frac{\pi{\omega_0}}{\kappa_r}\big)}{128M^2\pi{\omega_0}}
\sum_{l=0}^\infty(1+2l)\bigg[-|\,B_l({\omega_0}\sqrt{g_{00}})\,|^2
\\&+e^{2\pi{\omega_0}/\kappa_r}|\,B_l(-{\omega_0}\sqrt{g_{00}})\,|^2
\bigg]\;,\\
 C_H&=\mathcal{G}_H(0)-A_H\;,&
 \end{aligned} \right.
 \end{eqnarray}
 %%%%%%%%%%%%%%%%
\begin{eqnarray}\label{xishu-h2}
r\rightarrow{\infty}:~\left\{ \begin{aligned}
 A_H&\approx\frac{\omega_0}{4\pi}\coth\big(\frac{\pi\omega_0}{\kappa}\big)+\frac{F^{(+)}(\omega_0)}{r^2}\;,\\
 B_H&\approx\frac{\omega_0}{4\pi}+\frac{F^{(-)}(\omega_0)}{r^2}\;,\\
 C_H&=\mathcal{G}_H(0)-A_H\;.&
 \end{aligned} \right.
 \end{eqnarray}
Therefore, the probability of spontaneous transition and the
spontaneous excitation rate from ground state
 to exited state are respectively
\begin{equation}\label{ph1}
\mathcal{P}^H_{i\rightarrow{f}}({\tau})\approx\left\{
\begin{aligned}
  &\frac{1-e^{-\frac{\tau\omega_0}{\pi}\coth(\pi\omega_0/\kappa_r)}}{1+e^{2\pi\omega_0/\kappa_r}}\;,&r\rightarrow2M\;,
\\
&\frac{1-e^{-\frac{\tau\omega_0}{\pi}\coth(\pi\omega_0/\kappa)}}{1+e^{2\pi\omega_0/\kappa}}
\;,&r\rightarrow\infty\;.
\end{aligned} \right.
\end{equation}
and
\begin{equation}\label{gh1}
\Gamma_{i\rightarrow{f}}^H\approx\left\{
\begin{aligned}
  &\frac{{\omega_0}}{\pi(e^{2\pi{\omega_0}/\kappa_r}-1)}
  +\frac{f(\omega_0,2M)\,\omega_0}{\pi (e^{2\pi{\omega_0}/\kappa_r}-1)}\;,&r\rightarrow2M\;,
\\
&\frac{\omega_0}{\pi(e^{2\pi\omega_0/\kappa}-1)}+
\frac{f(-\omega_0,r)\,\omega_0}{\pi(e^{2\pi{\omega_0}/\kappa}-1)}\;,&r\rightarrow\infty\;.
\end{aligned} \right.
\end{equation}
%%%%%%%%%%%%%%%%%%%%%%%%%%%%%%%%%%%%%%%%%
%%%%%%%%%%%%%%%%%%%%%%%%%%%%%%%%%%%%%%%%%%
Here, one can see that the spontaneous excitation rate consists of
two pieces. One is the thermal radiation  outgoing from the
horizon (the first term in $\Gamma_{i\rightarrow{f}}^H$ when
$r\rightarrow2M$), which is identical to what we have obtained in
the Unruh case, and the other can be viewed as the thermal
radiation incoming from infinity (the first term in
$\Gamma_{i\rightarrow{f}}^H$ when $r\rightarrow\infty$). Both the
outgoing and incoming thermal radiation are backscattered by the
curvature on their way to a finite radial distance from the
horizon, and this backscattering is represented by the greybody
factors in the second term in $\Gamma_{i\rightarrow{f}}^H$ for
both cases. Therefore, one sees that the Hartle-Hawking vacuum
actually corresponds to a black hole immersed in a bath of both
incoming and outgoing thermal radiation. Again, this is consistent
with our understanding of the Hartle-Hawking vacuum gained in
other different contexts.

\subsection{The Boulware vacuum }
 Finally, let us briefly discuss the Boulware vacuum case. Now, the Wightman function can be written
 as \cite{wightman1,wightman2,wightman3}
 \begin{equation}
 G^+_B(x,x')=\sum_{lm}\int_{0}^{\infty}\frac{e^{-i\omega \Delta{t}}}{4\pi\omega}|\,Y_{lm}(\theta,\phi)\,|^2\big[|\,\overrightarrow{R_l}(\omega,r)\,|^2
+|\,\overleftarrow{R_l}(\omega,r)\,|^2\big]d\omega\;,
 \end{equation}
The corresponding Fourier transform reads
\begin{eqnarray}
\mathcal{G}_B({\lambda})=&&\int^{\infty}_{-\infty}e^{i\lambda{\tau}}{G_{B}}^+[x(\tau)]d\tau\nonumber\\
&&=\sum_{ml}\int^{\infty}_{0}\frac{d\omega}{4
\pi\omega}\int^{\infty}_{-\infty}e^{i({\lambda}-\omega/\sqrt{1-2M/r})\tau}|\,Y_{lm}(\theta,\phi)\,|^2\big[|\,\overrightarrow{R_l}(\omega,r)\,|^2
+|\,\overleftarrow{R_l}(\omega,r)\,|^2\big]d\tau\nonumber\\
&&=\sum_{ml}\int^{\infty}_{0}\frac{d\omega}{2\omega}\delta({\lambda}-\omega/\sqrt{1-2M/r})|\,Y_{lm}(\theta,\phi)\,|^2
\big[|\,\overrightarrow{R_l}(\omega,r)\,|^2
+|\,\overleftarrow{R_l}(\omega,r)\,|^2\big]\;.\label{b-green}
\end{eqnarray}
The further calculation should proceed in three separate cases
\begin{equation}\label{b-jifen}
\mathcal{G}_B({\lambda})=\left\{ \begin{aligned}
  &\sum_{ml}\frac{1}{2{\lambda}}|Y_{lm}(\theta,\phi)|^2\big[|\,\overrightarrow{R_l}({\lambda}\sqrt{g_{00}},r)\,|^2
+|\,\overleftarrow{R_l}({\lambda}\sqrt{g_{00}},r)\,|^2\big]\;,~&{\lambda}>0\;,
\\
&\sum_{ml}\frac{1}{4{\lambda}}|\,Y_{lm}(\theta,\phi)\,|^2\big[|\,\overrightarrow{R_l}({\lambda}\sqrt{g_{00}},r)\,|^2
+|\,\overleftarrow{R_l}({\lambda}\sqrt{g_{00}},r)\,|^2\big]\bigg|_{\lambda\rightarrow0}\;,&{\lambda}=0\;,
\\&0 , & {\lambda}<0\;.
\end{aligned} \right.
\end{equation}
This indicates that for a positive $\omega_0$,
$\mathcal{G}_B(-{\omega_0})$ vanishes. Thus a substitution of
Eq.~(\ref{b-jifen}) into Eq.~(\ref{ABC}) yields that $A_B=B_B\;$,
which in turn gives rise to $
\mathcal{P}_{i\rightarrow{f}}({\tau})=0\; $ and  $
\Gamma_{i\rightarrow{f}}=2A_B-2B_B=0\;$. Therefore, no spontaneous
excitation would ever occur in the Boulware vacuum. As a result, the
Boulware vacuum corresponds to our familiar notion of a vacuum
state.

%%%%%%%%%%%%%%%%%%%%%%%%%%%%%%%%%%%%%

\section{Conclusion}

In summary, we have examined the Hawking radiation from the point of
view of open quantum systems by looking at the time evolution of a
detector (modelled by a two-level atom) interacting with vacuum
massless scalar fields. The time evolution of the detector is
governed by a master equation obtained by tracing over the field
degrees of freedom from the complete system. The nonunitary effects
have been studied by analyzing the time behavior of the detector's
admissible state in the Unruh, Hartle-Hawking, as well as Boulware
vacua outside a Schwarzschild black hole. It is found that the
detector in the Unruh and Hartle-Hawking vacua would spontaneously
excite with  a nonvanishing probability the same as what one would
obtain if there is thermal radiation at the Hawking temperature from
the black hole, reproducing the basic results concerning the Hawking
effect in the framework of open quantum systems. However, as we have
pointed out in Section II, the possibility of a nonvanishing
spontaneous excitation rate in fact involves the phenomena of
decoherence and disspation, so the open system approach seems more
comprehensive physically than traditional treatments. Our study
suggests that the general techniques and results in the theory of
open quantum systems not only is applicable to the study of the
Hawking effect and possibly even to that of other phenomena in
curved space-times, such as particle creation, but also may shed new
light on the physical understanding of them.  We hope to turn to
these issues in the future.

\begin{acknowledgments}
 One of us (HY) would like to thank C.P. Sun for interesting
 discussions, R.G. Cai and the Kavli Institute for Theoretical Physics China,
where part of the work was done
 during the String Theory and Cosmology Program, for warm hospitality,
 This work was supported in part by the National
Natural Science Foundation of China  under Grants No. 10575035 and
 No. 10775050, and the Program for NCET under Grant No. 04-0784.
\end{acknowledgments}

\end{document}